\documentclass[conference,compsoc]{IEEEtran}
\ifCLASSINFOpdf
\else
\fi
\usepackage{cite}
\usepackage{amsmath,amssymb,amsfonts}
\usepackage{graphicx}
\usepackage{textcomp}
\usepackage{xcolor}
\usepackage{gensymb} 
\usepackage{algorithm}
\usepackage{algorithmic}
\usepackage[multiple]{footmisc}

\usepackage[shortlabels]{enumitem}

\usepackage{comment} 
\usepackage{float} 
\usepackage{longtable} 
\usepackage{enumitem} 

\begin{document}
%

\title{Adaptive Priority-based Conflict Resolution of IoT Services \vspace{-4mm}}

\author{Dipankar Chaki and Athman Bouguettaya\\
\textit{School of Computer Science, University of Sydney, Australia}\\
\textit{Email: \{dipankar.chaki, athman.bouguettaya\}@sydney.edu.au}\vspace{-4mm}
}

\maketitle

\IEEEoverridecommandlockouts
\IEEEpubid{\begin{minipage}{\textwidth}\ \\[12pt] \centering
  \copyright 20XX IEEE.  Personal use of this material is permitted.  Permission from IEEE must be obtained for all other uses, in any current or future media, including reprinting/republishing this material for advertising or promotional purposes, creating new collective works, for resale or redistribution to servers or lists, or reuse of any copyrighted component of this work in other works.
\end{minipage}}

\begin{abstract}
We propose a novel conflict resolution framework for IoT services in multi-resident smart homes. An adaptive priority model is developed considering the residents' contextual factors (e.g., age, illness, impairment). The proposed priority model is designed using the concept of the analytic hierarchy process. A set of experiments on real-world datasets are conducted to show the efficiency of the proposed approach.
\vspace{-2mm}
\end{abstract}

\vspace{-2mm}

\begin{IEEEkeywords}
IoT service; Multi-resident smart home; Adaptive priority; Analytic hierarchy process; Conflict resolution\vspace{-6mm}
\end{IEEEkeywords}

%
\IEEEpeerreviewmaketitle

\section{Introduction}
\vspace{-2mm}
Internet of Things (IoT) refers to the billions of physical objects (a.k.a. things) that are connected to the Internet \cite{nauman2020multimedia}. These ``things" are evolving due to the convergence of multiple technologies, commodity sensors, machine learning, real-time analytics, ubiquitous computing, and embedded systems \cite{marwedel2021embedded}. IoT technology is the key ingredient for many cutting-edge applications such as smart campuses, smart cities, intelligent transport systems, and smart grids. One of the prominent application domains of IoT is \emph{Smart homes}. A smart home is any regular home that is equipped with IoT devices. These IoT devices monitor the usage patterns of everyday ``things". The purpose of a smart home is to provide its residents with \textit{convenience} and \textit{efficiency} \cite{huang2018convenience}.

Things in the IoT environment exhibit the same behavior as represented in the \emph{service paradigm} \cite{huang2016service}. Each ``thing" has a set of \emph{functional} and \emph{non-functional} properties. This is why the service paradigm is leveraged as a framework to define the functional and non-functional properties of smart home devices as \emph{IoT services} \cite{huang2018discovering}. For instance, a light bulb in a smart home is regarded as a light service. The functional property of the light service is to provide illumination. Examples of non-functional properties include luminous intensity, color (e.g., warm white), connectivity.


Different residents may have different preferences (a.k.a., service requirements) in a multi-occupant environment that may lead to \emph{IoT service conflicts} \cite{chaki2020fine}. For instance, one resident may prefer 25\degree C AC temperature, while another resident may prefer 20\degree C. Therefore, a service conflict occurs since the AC service cannot satisfy multiple residents' requirements at the same time and location. In this regard, \textit{detecting} and \textit{resolving} conflicts has become essential to make the residents' lives convenient and efficient.


Existing research mainly adopts three strategies for conflict resolution: (i) \emph{Fair Principle}, (ii) \emph{Use First}, and (iii) \emph{Priority Assignment} \cite{nurgaliyev2017improved,lee2019situation,lalanda2017conflict,lakhdari2020fluid,abusafia2020reliability}. The fair principle strategy does not prioritize any resident. Instead, it calculates the average preference values. For example, two residents (R1, R2) are in the living room, and R1 prefers 24\degree C temperature while R2 prefers 20\degree C. The fair principle method sets the indoor temperature to the average of these two values (i.e., 22\degree C). Use first is another technique for conflict resolution and this method only considers the user's settings, whoever starts using it first. Another widely used approach for conflict resolution is pre-defined user priority, e.g., in some cultures, parents have higher priority than children.


We identify a few shortcomings of the existing approaches. The existing frameworks do not consider contextual factors such as age, role, impairment (e.g., visual impairment, hearing impairment), and illness (e.g., cold, fever) when assigning priorities to the residents. For example, resident R1 has a cold and prefers 25\degree C temperature, while resident R2 does not have any cold and prefers 19\degree C. In this context, the framework may give up the fair principle approach. The framework may favor resident R1 and set the temperature close to R1's preference. Furthermore, the priority might not be static. For example, R2 has a hearing impairment and a conflict related to the TV volume occurs with R1. Now, resident R2 will be prioritized. Hence, priority assignment is context-aware and complex. There is a need for an adaptive priority assignment technique based on the above-mentioned contextual factors. This adaptive priority assignment scheme provides the baseline for conflict resolution to enhance the residents' overall satisfaction.

We propose a novel approach for conflict resolution by assigning priority based on contextual factors. We identify four factors: age, illness, visual impairment, and hearing impairment to build the adaptive priority model. Priority allocation is challenging since factors are time-dependent. Some factors are transient, and some are permanent. For example, illness is categorized as a transient factor as people may temporarily catch a cold/fever, and visual impairment can be considered a permanent factor. We design the priority model using a multi-criteria decision-making approach, Analytic Hierarchy Process (AHP) \cite{saaty1994make}. We use AHP as it has been applied in other areas as a state-of-art prioritizing methodology. Our goal is to prioritize residents based on contextual criteria such as age, illness, visual impairment, and we score each resident against these criteria. Our approach adapts the weight of each criterion based on dynamic changes in contexts. The adaptation is achieved using consistency ratio as an indicator to update the weight matrix. This step increases the consistency of the weight assignment, which enhances the accuracy of residents’ priority ranking (details are illustrated in section 4.2). The contributions of this paper are summarized as follows:


\begin{itemize}[leftmargin=*]
    \item An adaptive priority allocation strategy using the AHP concept that assigns priority based on contextual factors. 
    \item Experimental evaluation is conducted to demonstrate the effectiveness of the proposed approach.
\end{itemize}

\vspace{-3mm}
\section{Motivation Scenario}
\vspace{-2mm}

We discuss the following two scenarios to illustrate the notion of adaptive priority allocation to the users based on contextual factors. We consider four contextual factors: resident's age, their visual impairment (VI) status, hearing impairment (HI) status, and illness (e.g., cold/fever).

\begin{figure}[htbp]
\vspace{-3mm}
\center
\includegraphics[width=\columnwidth,height=5.5cm,keepaspectratio]{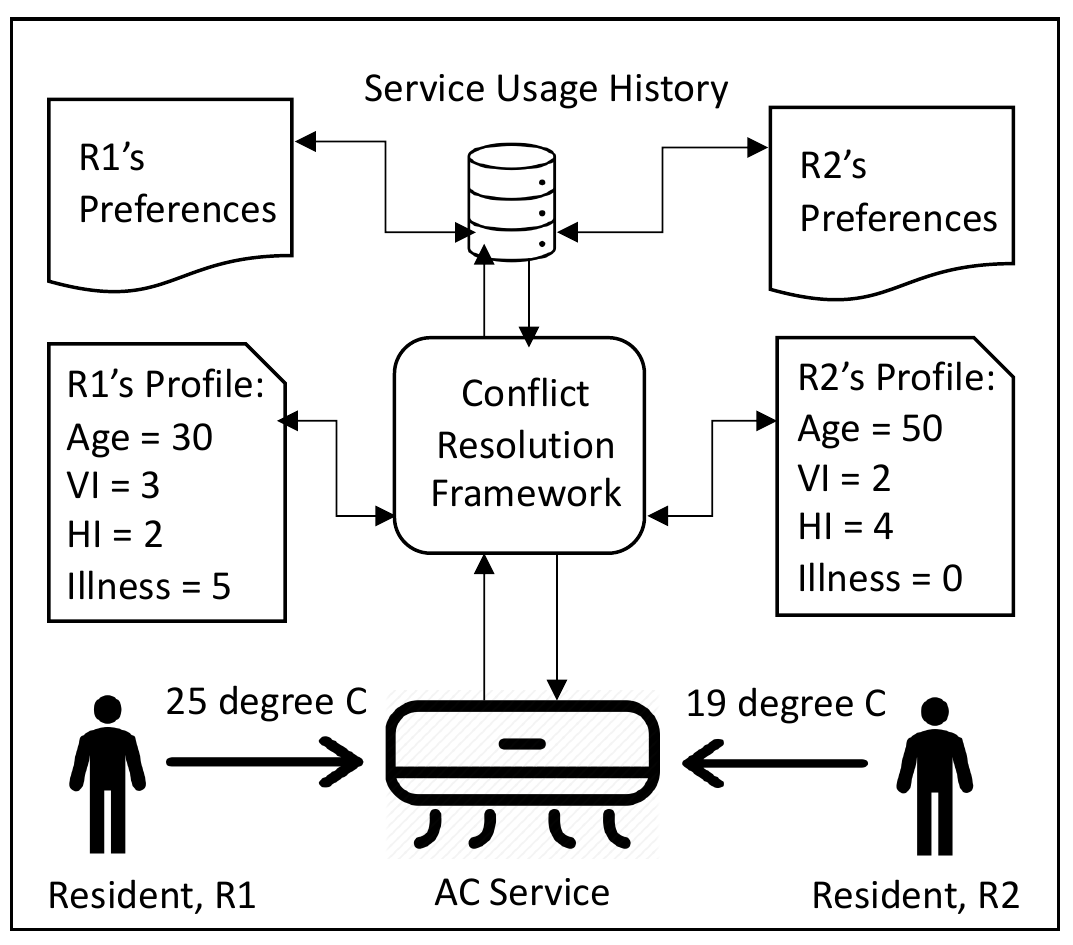}
\vspace{-4mm}
\caption{Priority allocation based on illness status.}
\label{motivation1}
\vspace{-2mm}
\end{figure}

\textit{Scenario 1:}
Suppose two residents (R1, R2) are trying to use an AC service simultaneously at the same location. R1's requirement is 25\degree C and R2's requirement is 19\degree C (Fig. \ref{motivation1}). A conflict occurs since the AC service cannot satisfy the requirements of these two residents simultaneously. The simple approach to resolving this conflict calculates the average value of these two requirements and sets the AC temperature to 22\degree C. Another approach to resolve this conflict is to assign priority to one resident and set the AC temperature as per their preference. Age can be used to assign priority. Older people will have higher priority than younger people. According to the profiles of the residents from Fig. \ref{motivation1}, R2 ($R2_{Age} = 50$) will have higher priority since they are older than R1 ($R1_{Age} = 30$). In this case, R1 will be highly unsatisfied, and the pre-defined priority allocation is the main drawback of this approach. Note that R1 has caught a cold/fever and is sicker than R2 ($R1_{Illness} > R2_{Illness}$). The higher the value, the more ill/sick the person is. R1 may need to be prioritized in this context, and the AC temperature may be set close to R1's preference (i.e., 23-24\degree C).

\begin{figure}[htbp]
\vspace{-3mm}
\center
\includegraphics[width=\columnwidth,height=5.5cm,keepaspectratio]{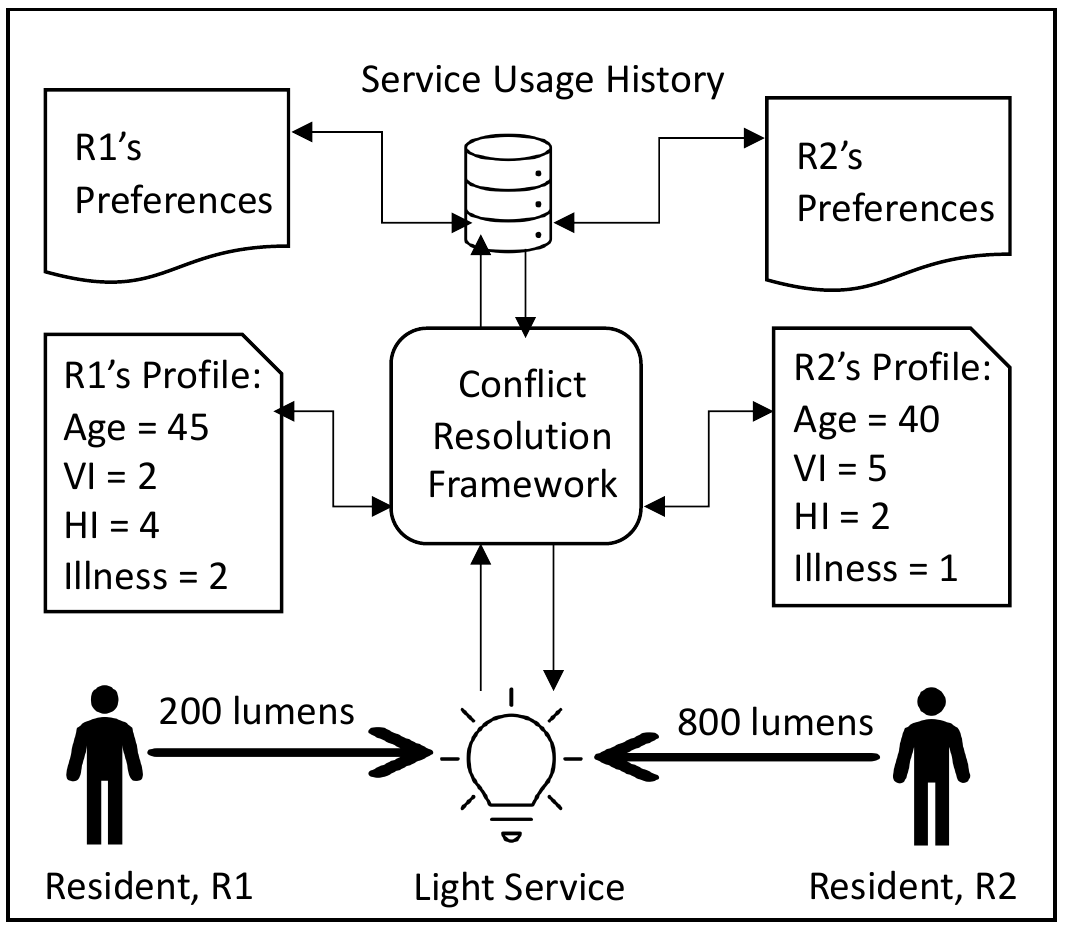}
\vspace{-4mm}
\caption{Priority allocation based on visual impairment status.}
\label{motivation2}
\vspace{-4mm}
\end{figure}

\textit{Scenario 2}:
Suppose two residents (R1, R2) are trying to use a light service simultaneously at the same location. R1's requirement is 200 lumens and R2's requirement is 800 lumens (Fig. \ref{motivation2}). A conflict occurs since the light service cannot satisfy the requirements of these two residents simultaneously. The simple approach to resolving this conflict calculates the average value of these two requirements and sets the light luminosity at 500 lumens. If age is used to assign priority to one resident, then R1 ($R1_{Age} = 45$) will have higher priority since they are older than R2 ($R2_{Age} = 40$). However, it can be seen that R2 is visually impaired (VI). Visual impairment is regarded as a decreased ability to see that is not fixable by glasses or contact lenses. To some extent, those who have a decreased ability to see due to not having access to glasses or contact lenses are also regarded as visually impaired. In this research, a higher VI score denotes a high vision impairment state. R2 may need to be prioritized in this circumstance, and the light illumination may be set close to R2's preference (i.e., 600-700 lumens).

Therefore, priority assignment is context-specific, complex, and dynamic. There is a need for an adaptive priority assignment technique based on the contextual factors which is essential for conflict resolution. 


\vspace{-3mm}
\section{Preliminaries}
\vspace{-2mm}
We first define \textit{IoT service} and \textit{IoT service event} to explain the \emph{IoT service conflict} concept. The definitions of IoT service and IoT service event are acquired from \cite{chaki2020conflict}. 

An \textit{IoT Service ($S$)}, is a tuple of \big \langle \textit{$S_{id}$, $S_{name}$, $F$, $Q$}\big \rangle.

\begin{itemize}[itemsep=0ex, leftmargin=*]
  \item $S_{id}$ represents the unique service identifier (ID).
  \item $S_{name}$ is the name of the service.
  \item $F$ is a set of \big \{\textit{$f_1$, $f_2$, $f_3$,.......$f_n$}\big \} where each $f_i$ is a functional attribute of a service. 
  \item $Q$ is the set of \big \{\textit{$q_1$, $q_2$, $q_3$,.......$q_m$}\big \} where each $q_j$ is a non-functional attribute of a service.
\end{itemize} 


An \textit{IoT Service Event ($SE$)}, is an instantiation of a service. When a service manifestation occurs (i.e., turn on, turn off, increase, decrease, open, close), an event records the service state along with its user, execution time and location. A service event is a tuple of \big \langle \textit{\{$S_{id}, F, Q\}, T, L, U$}\big \rangle.

\begin{itemize}[itemsep=0ex, leftmargin=*]
  \item $S_{id}$ is a unique service ID.
  \item $F$ is a set of functional attributes.
  \item $Q$ is a set of non-functional attributes.
  \item $T$ is a set of \{\textit{$T_s$,$T_e$}\} where $T_s$ and $T_e$ represent the service start time and end time, respectively.
  \item $L$ is the service location.
  \item $U$ is the service user.
\end{itemize} 


An IoT service ($S$) is associated with a set of functional and non-functional properties. An IoT service event illustrates a resident's particular service usage in conjunction with time and location. IoT service event sequences ($SES$) record all the history of service events. A conflict may emerge due to different residents' different service requirements. Consequently, a conflict resolution ($Res$) technique is required to maximize the satisfaction of the residents. Given these pieces of information, the paper aims to identify a function $F(S, SES)$, where $Res \approx F(S, SES)$. 

\vspace{-2mm}

\section{Conflict Resolution Framework}
\begin{figure}[htbp]
\center
\includegraphics[width=\columnwidth]{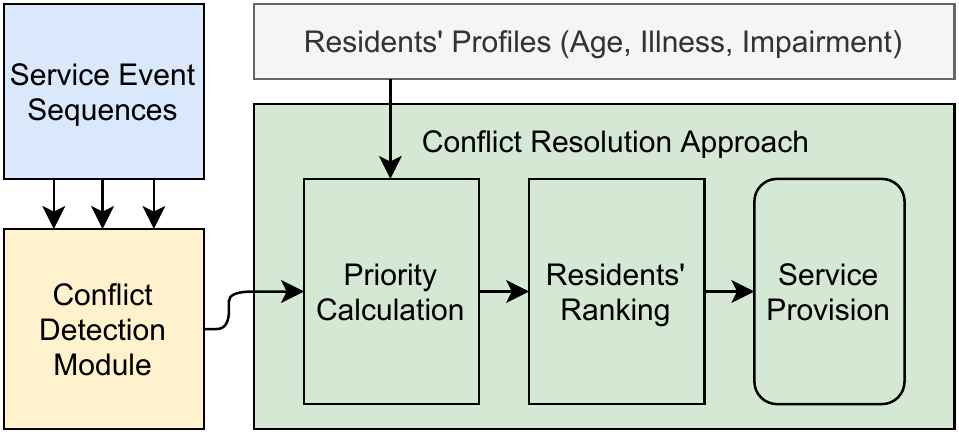}
\vspace{-6mm}
\caption{IoT service conflict resolution framework.} 
\label{framework}
\vspace{-4mm}
\end{figure}

\vspace{-2mm}
The proposed conflict resolution framework has mainly 2 modules: (i) conflict detection module, and (ii) conflict resolution approach (Fig. \ref{framework}).

\vspace{-2mm}

\vspace{-1mm}


\vspace{-1mm}

\subsection{Conflict Detection Module}
\vspace{-1mm}
An IoT service conflict occurs when a service cannot satisfy the requirements of multiple users at the same time and location. Conflicts are defined considering the requirements of occupants, and these requirements are generated from the IoT service events. Given two service events ($SE_i$, $SE_j$), the following conditions have to be satisfied to be considered as a conflict situation.
\begin{itemize}[itemsep=0ex, leftmargin=*]
    \item $L_{S_i}$ $\simeq$ $L_{S_j}$, meaning that two services ($S_i$, $S_j$) are executed at the same location.
    \item $(T_{s_i}, T_{e_i}) \cap (T_{s_j}, T_{e_j})) \neq \emptyset$, denoting that two service events ($SE_i$, $SE_j$) are invoked at the same time and there is a temporal overlap between them.
    \item $U_{S_i} \neq U_{S_j}$, meaning that these two events are invoked by two different users.
    \item $\exists Q_k \in S.Q: S_i.Q_k \neq S_j.Q_k$; there exist at least one property which is different between $S_i.Q$ and $S_j.Q$.
\end{itemize}

We adopt the conflict detection algorithm proposed in \cite{chaki2020conflict}. This module is not the core of our contributions; however, it produces the input for the conflict resolution module, which holds the present work's core contributions.

\vspace{-3mm}

\subsection{Conflict Resolution Approach}


\vspace{-2mm}

The output of the conflict detection module is the name of the conflicting services, type of conflict, and the overlapping time-period where a conflict occurs, which are the input of this module. Also, the profiles of the residents are the input of this module. Each profile contains each resident's current age, illness status, and impairment status (i.e., hearing impairment level and visual impairment level). We use Analytic Hierarchy Process (AHP) to assign priority to the residents based on their profile data.

AHP proposed by Saaty is very popular and has been applied in various areas, including planning, selecting the best alternative, resource allocation, and resolving conflict \cite{saaty1994make}. AHP decomposes the multi-criteria decision-making problem into a hierarchical structure consists of a goal, criteria, and decision elements (a.k.a., alternatives). A pairwise comparison between alternatives is conducted at each level in the hierarchical structure. AHP works by assigning global and local priorities (normalized weights) to the criteria and alternatives, respectively, according to their importance. These global and local priorities create a pairwise comparison matrix that constitutes the decision ranking for the alternatives contributing towards the goal. A pairwise comparison among alternatives is made using the 1–9 scale described in Table \ref{ahp scale}. We perform the following five steps:

\vspace{-2mm}
\begin{table}[htbp]
\caption{AHP Scale for Pairwise Criteria Comparison}
\vspace{-2mm}
\label{ahp scale}
\centering
\begin{tabular}{|c|c|p{6.3cm}}
\hline
\textbf{Scale} & \textbf{Interpretation}\\
\hline
1 & Equal importance\\
\hline
3 & Moderate importance of one over another \\
\hline
5 & Essential or strong importance\\
\hline
7 & Very strong importance \\
\hline
9 & Extreme importance \\
\hline
2, 4, 6, 8 & Intermediate values between two adjacent criteria \\
\hline
Reciprocals & \parbox{6.3cm}{If criterion $x$ has one of the above numbers assigned to it when compared with criterion $y$, then $y$ has the reciprocal value when compared with $x$}\\
\hline
\end{tabular}
\vspace{-3mm}
\end{table}

\begin{enumerate}[itemsep=0ex, leftmargin=*]
  \item The hierarchical structure is constructed, placing the goal at the top. The criteria, and sub-criteria (if any) are positioned at the subsequent levels underneath the goal. Our goal is to rank residents (i.e., prioritization) based on contextual criteria (i.e., age, illness, visual impairment and hearing impairment).
  \item The pairwise comparison matrix (A) is built using the scale introduced in Table \ref{ahp scale} for the n decision elements at each level of the hierarchy, as shown in Equation \eqref{eq2}.
  
  \vspace{-2mm}
  \begin{equation}
  \label{eq1}
  A = 
      \begin{bmatrix}
        a_{11} & a_{12} & a_{13} & ... & a_{1n}\\
        a_{21} & a_{22} & a_{23} & ... & a_{2n}\\
        a_{31} & a_{32} & a_{33} & ... & a_{3n}\\
        ... & ... & ... & ... & ...\\
        a_{n1} & a_{n2} & a_{n3} & ... & a_{nn}\\
      \end{bmatrix}
  \end{equation}
  where $a_{ij}$ represents a pairwise comparison and satisfies the following condition:
  \begin{equation*}
      a_{ij} = 
      \begin{cases}
      1,& \text{if } i = j \\
      1/a_{ji}, & \text{if } i \neq j
      \end{cases}
      \text{ and } a_{ij} > 0
  \end{equation*}
  \vspace{-2mm}
  Therefore, the comparison matrix can be written as:
  \begin{equation}
  \label{eq2}
  A = 
      \begin{bmatrix}
        1 & a_{12} & a_{13} & ... & a_{1n}\\
        1/a_{12} & 1 & a_{23} & ... & a_{2n}\\
        1/a_{13} & 1/a_{23} & 1 & ... & a_{3n}\\
        ... & ... & ... & ... & ...\\
        1/a_{1n} & 1/a_{2n} & 1/a_{3n} & ... & 1\\
      \end{bmatrix}
  \end{equation}
  \vspace{-2mm}
  
We demonstrate the ranking procedure considering the example from motivation scenario 1. We build the pairwise comparison matrix (Table \ref{pairwise matrix}) using equation \eqref{eq2} based on the criteria shown in Fig. 1. Since the conflict occurs in AC temperature, illness criterion is given high priority. In order to assign priority, we follow the scale mentioned in Table \ref{ahp scale}. This priority assignment will change based on contextual factors. For example, in scenario 2, high priority will be given to visual impairment criterion since the conflict occurs in light illumination.
  
\vspace{-2mm}  
\begin{table}[htbp]
\caption{Pairwise Criteria Matrix}
 \vspace{-2mm}
\label{pairwise matrix}
\centering
\begin{tabular}{|c|c|c|c|c|}
\hline
 & Age & VI & HI & Illness\\
\hline
Age & 1 & 1/3 & 1/3 & 1/7\\
\hline
VI & 3 & 1 & 1 & 1/5\\
\hline
HI & 3 & 1 & 1 & 1/5\\
\hline
Illness & 7 & 5 & 5 & 1 \\
\hline
\end{tabular}
\vspace{-6mm}
\end{table}
  
After creating the pairwise comparison matrix, a vector of weights $W = [W_1, W_2, W_3, ... W_n]$ is populated based on Saaty’s eigenvector procedure. There are two steps to calculate the weights: normalizing the pairwise matrix and creating a weighted matrix. The comparison matrix is normalized using Equation \eqref{eq3} and create a vector $X = [X_1, X_2, X_3, ... X_n]$ \cite{chen2006applying}.
  
\vspace{-2mm}
 \begin{equation}
  \label{eq3}
      X_k = {\left (\prod_{j=1}^{n}A_{ij} \right )}^{(1/n)}
  \end{equation}
  \vspace{-3mm}
  
We apply Equation \eqref{eq3} to our running example and get the value of $X$ for each criterion as:

\vspace{-3mm} 
\begin{table}[htbp]
\caption{Normalized Values for Criteria}
\vspace{-2mm}
\label{normalized matrix}
\centering
\begin{tabular}{|c|c|c|c|c|}
\hline
 Age & VI & HI & Illness & Sum\\
\hline
0.355 & 0.880 & 0.880 & 3.637 & 5.752\\
\hline
\end{tabular}
\vspace{-3mm}
\end{table}
  
\item The comparison matrix produces the priority vector with relative weights to form the ranking considering the importance of the decision elements. The normalized principal eigenvector of matrix A generates the priority vector. The n × n pairwise comparisons are consolidated into n measures of intensity by the eigenvector. This step is repetitive for each criterion. Equation \eqref{eq4} computes the weights for each of the element. To  obtain the weighted matrix, number of element ($n$) is divided by the sum of the column of the matrix \cite{hasan2018novel}.
  
  \vspace{-2mm}
  \begin{equation}
      \label{eq4}
      W_l=\frac{X_k}{\sum_{k=1}^{n}X_k}
  \end{equation}
  \vspace{-2mm}
  
Using Equation \eqref{eq4}, we get the weight matrix as follows:
 
\vspace{-3mm} 
\begin{table}[htbp]
\caption{Weights for Criteria}
\vspace{-2mm}
\label{weight matrix}
\centering
\begin{tabular}{|c|c|c|c|}
\hline
 Age & VI & HI & Illness\\
\hline
0.617 & 0.153 & 0.153 & 0.632\\
\hline
\end{tabular}
\vspace{-3mm}
\end{table}
  
\item The pairwise comparison matrix ($A$) is multiplied by the transpose of weight matrix ($W^T$) to get the weighted sum vector ($S$) using Equation \eqref{eq5}.
  
  \vspace{-3mm}
  \begin{equation}
  \label{eq5}
    \begin{bmatrix}
        s_{1}\\
        s_{2}\\
        s_{3}\\
        ... \\
        s_{n}\\
      \end{bmatrix}
      =
    \begin{bmatrix}
        1 & a_{12}  & ... & a_{1n}\\
        1/a_{12} & 1  & ... & a_{2n}\\
        1/a_{13} & 1/a_{23} & ... & a_{3n}\\
        ... & ... & ... & ...\\
        1/a_{1n} & 1/a_{2n} & ... & 1\\
      \end{bmatrix}
      *
      \begin{bmatrix}
        W_{1}\\
        W_{2}\\
        W_{3}\\
        ... \\
        W_{n}\\
      \end{bmatrix}
  \end{equation}
  
 \vspace{-2mm}
  
Then, the weighted sum vector is divided by the weight vector to get the consistency vector ($CV$) (Equation \eqref{eq6}). 
  
  \vspace{-2mm}
  \begin{equation}
  \label{eq6}
      CV_n = 
      \begin{bmatrix}
        s_{1}\\
        s_{2}\\
        ... \\
        s_{n}\\
      \end{bmatrix}
      /
      \begin{bmatrix}
        W_{1}\\
        W_{2}\\
        ... \\
        W_{n}\\
      \end{bmatrix}
  \end{equation}
  \vspace{-2mm}
  
  $CV$ is shown for the running example in Table \ref{consistency vector}.
  
  \vspace{-3mm}
  \begin{table}[htbp]
\caption{Consistency Vector for Criteria}
 \vspace{-2mm}
\label{consistency vector}
\centering
\begin{tabular}{|c|c|c|c|}
\hline
 Age & VI & HI & Illness\\
\hline
4.11688 & 4.03641 & 4.03641 & 4.10292\\
\hline
\end{tabular}
\vspace{-3mm}
\end{table}
  
There is a relationship between the weight vector ($W$) and the comparison matrix ($A$). This relationship is shown in Equation \eqref{eq7}.
  
  \vspace{-3mm}
  \begin{equation}
      \label{eq7}
      A_W = \lambda_{max}W
  \end{equation}
  \vspace{-4mm}
  
  where $\lambda_{max}$ is the average of the consistency vector's values and it is used to calculate the consistency index ($CI$). $\lambda$ is obtained using the following Equation \eqref{eq8}. We get $\lambda = 4.07315$ for the running example.
    
    \vspace{-2mm}  
    \begin{equation}
      \label{eq8}
      \lambda=\frac{\sum_{i=1}^{n}CV_i}{n}
    \end{equation}
    \vspace{-3mm}
  
  The consistency index ($CI$) for each matrix is calculated by Equation \eqref{eq9}. We get $CI = 0.02438$ for our example.
  
  \vspace{-2mm}
  \begin{equation}
  \label{eq9}
      CI = \frac{\lambda_{max}-n}{n-1}
  \end{equation}
  \vspace{-3mm}
  
  The  consistency  ratio ($CR$) is  the  ratio  of $CI$ and $RI$ as shown in Equation \eqref{eq10}. We get, $CR = 0.02740$.
  
  \vspace{-2mm}
  \begin{equation}
  \label{eq10}
      CR = \frac{CI}{RI}
  \end{equation}
  \vspace{-3mm}
  
  where $RI$ is called Random Index. The values of $RI$ for n = 1 to 10 are displayed in Table \ref{random index}. If $CR$ is greater than 0.1, the pairwise comparisons is needed to be revised. If it is less than or equal to 0.1, the inconsistency is acceptable. In our case, $CR$ is less than 0.1. Therefore, the weight we assign at the beginning is acceptable.
  
 \vspace{-3mm}
\begin{table}[htbp]
\caption{Value of Random Index (RI)}
 \vspace{-2mm}
\label{random index}
\centering

\begin{tabular}{|c|c|c|c|c|c|c|c|c|c|}
\hline
\textbf{n} & 1 & 2 & 3 & 4 & 5 & 6 & 7 & 8 & 9\\
\hline
\textbf{RI} & 0 & 0 & 0.52 & 0.9 & 1.12 & 1.24 & 1.32 & 1.41 & 1.45\\
\hline
\end{tabular}
\end{table}

    \item Individual evaluation is averaged using Equation \eqref{eq11} to get the consolidated measure of the pairwise comparisons that are involved in decision making.
    
    \vspace{-2mm}
    \begin{equation}
        \label{eq11}
        a_{ij}^{hp} = \sqrt[q]{\prod_{q=1}^q a_{ij}^q}
    \end{equation}
    \vspace{-2mm}
    
    where $a_{ij}^q$ is an element of matrix $A$ of a particular decision maker and $a_{ij}^{hp}$ is the mean of all the involved decision makers. Now consider the following individual data collected from motivating scenario 1 to rank the residents (Table \ref{individual}). From this data, using Equation \eqref{eq11}, we get the weight and ranking of the residents (Table \ref{rank}). We then normalize the weight value and use it for the service provision.
    
\vspace{-3mm}
\begin{table}[htbp]
\caption{Individual Data of the Residents}
 \vspace{-2mm}
\label{individual}
\centering
\begin{tabular}{|c|c|c|c|c|}
\hline
Resident & Age & VI & HI & Illness\\
\hline
R1 & 30 & 3 & 2 & 5\\
\hline
R2 & 50 & 2 & 4 & 0\\
\hline
\end{tabular}
 \vspace{-3mm}
\end{table}

 \vspace{-2mm}
\begin{table}[htbp]
\caption{Weight and Ranking of the Residents}
 \vspace{-2mm}
\label{rank}
\centering
\begin{tabular}{|c|c|c|c|c|}
\hline
Resident & Weight & Normalized Weight & Rank\\
\hline
R1 & 0.92349924 & 0.720727015 & 1\\
\hline
R2 & 0.357844765 & 0.279272985 & 2\\
\hline
\end{tabular}
\vspace{-1mm}
\end{table}
    
\end{enumerate}

We use the normalized weight value to set the AC temperature. For example, R1's requirement is 25\degree C, and R2's requirement is 19\degree C. The AC temperature will be set as $(25 \times 0.72) + (19 \times 0.28) = 23.32 \approx 24$. We tend to set the temperature leaning towards the higher priority resident.

\section{Experimental Results and Discussion}
\vspace{-2mm}
\subsection{Experimental Setup}
\vspace{-2mm}
We use 4 individual residents' activity data (labels HH102, HH104, HH105, HH106) from the CASAS dataset \cite{cook2012casas}. We select these labels as they contain activities of a similar period (between June 15, 2011, and August 14, 2011). We merge them to mimic the environment of multi-resident smart homes and deduce conflicting situations. The dataset consists of various sensors, such as light sensors, temperature sensors, magnetic door sensors, infrared motion sensors, and light switches. Each sensor is regarded as an IoT service. Descriptions of dataset attributes are displayed in Table \ref{data}. The dataset has light illumination values and temperature values with timestamps. We consider these values as the resident's preferred values for light and AC. The dataset does not have any profile information of any user. We augment the missing values by randomly assigning values to age, illness, visual impairment, and hearing impairment attributes based on a uniform distribution. 

\begin{table}[htbp]
\vspace{-3mm}
\caption{Description of the dataset attributes}
\vspace{-2mm}
\label{data}
\begin{tabular}{|c|l|p{6.3cm}}
\hline
\textbf{Attributes} &  \textbf{Description} \\
\hline
Date &  The service execution date\\ \hline
Time &  The service execution time\\ \hline
Sensor & \parbox{6.5cm}{Name of the sensors such as motion sensors, light switch, light sensors, door sensors, temperature sensors}\\\hline
Status & \parbox{6.5cm}{ON, when the service starts, and OFF, when the service stops}\\
\hline
\end{tabular}
\vspace{-3mm}
\end{table}

\vspace{-3mm}

\subsection{Performance Evaluation}
\vspace{-2mm}

The dataset has overlapping service event sequences and the residents' requirements. However, it does not have any ground truth value for evaluation. We create ground truth using Monte Carlo methods \cite{mooney1997monte}. The Monte Carlo simulation model predicts by using a range of values in the problem domain rather than a specific input. This method leverages probability distributions (normal, Gaussian, uniform) for any variable with uncertainty. The process of using random values is repeated numerous times based on the specified trial numbers \cite{fattah2019long}. Generally, the more trials, the higher likelihood the outcome will converge to a value. In this regard, we generate ground truth values using three different probability distributions such as normal distribution, uniform distribution, and triangular distribution. We then conduct three sets of experiments based on three different sets of ground truth values to show the efficiency and effectiveness of the first phase of the proposed conflict resolution framework.



\begin{figure}[htbp]
\begin{center}
\includegraphics[width=.85\columnwidth,height=5cm]{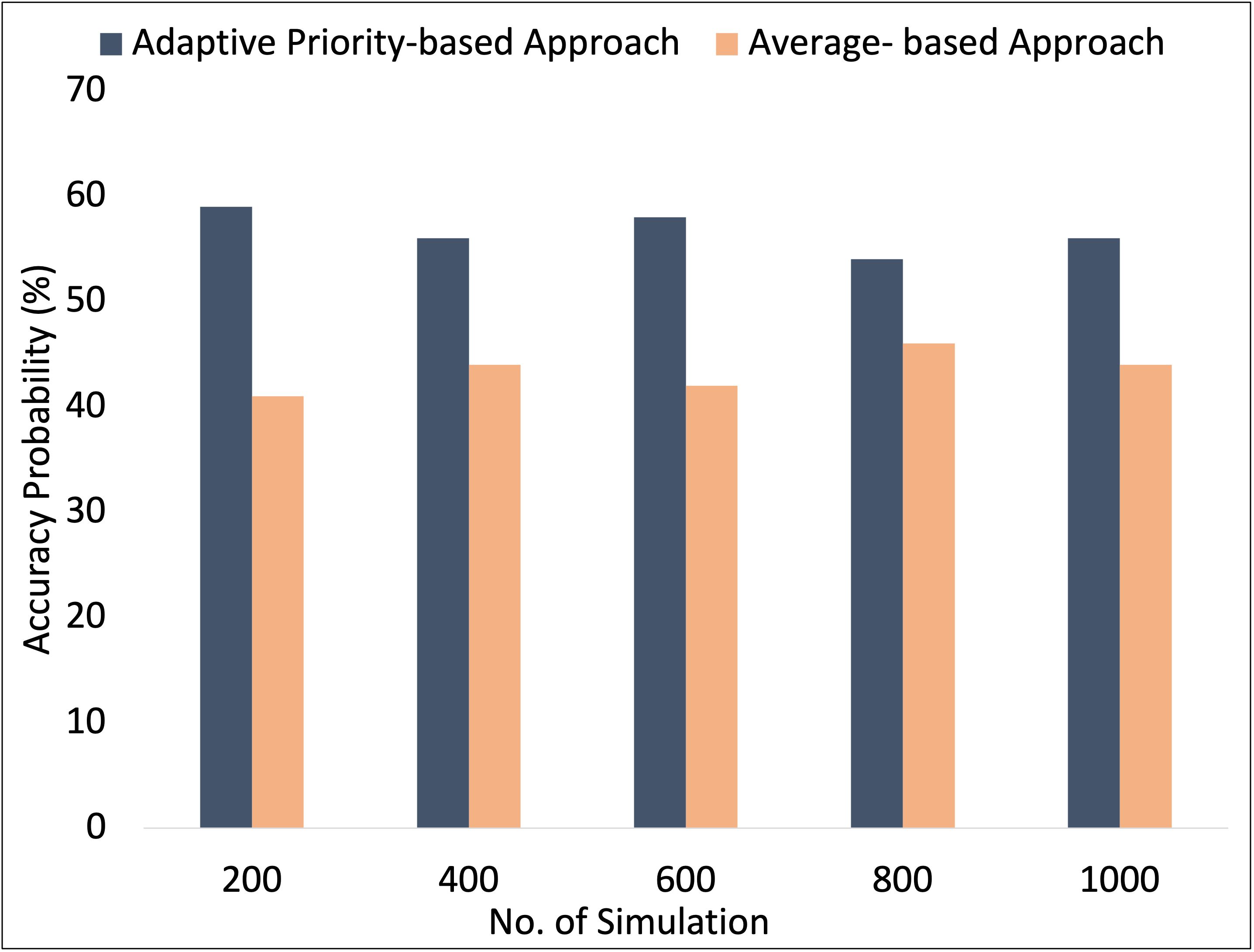}
\vspace{-2mm}
\caption{Accuracy probability in normal distribution.}
\vspace{-3mm}
\label{fig4}
\end{center}

\end{figure}

The first set of experiments is conducted considering ground truth values generated from a normal distribution. The results are shown in Fig. \ref{fig4}. We compare our approach (Adaptive Priority-based) with an existing approach (Average-based) proposed in \cite{ospan2018context}. At first, we identify the independent and dependent variables and define their domain of possible inputs—for example, light illumination, AC temperature based on the conflict type. Then, we determine a probability distribution to generate inputs over the domain randomly. We use previous history data to determine probability distribution. In the example mentioned in motivation scenario 1, the mean of R1's and R2's temperature requirement is 22\degree C. Since R1 is sick, we scan R2's previous AC usage data to find out their preferred temperature when they were sick. Based on R1's current requirement (25\degree C) and R2's previous history (23\degree C during sickness), we set the mean value as 24\degree C in the probability distribution function. After that, we compute the output based on the randomly generated inputs. Our goal is to estimate the output value close to the randomly generated ground-truth value based on the adaptive priority-based approach and the average-based approach. For example, both the average-based approach and adaptive priority-based approach propose a temperature value. Whichever value is close to the ground truth is considered as an accurate value for that particular conflict. We run the simulations 200 times, 400 times, 600 times, and so on. In total, we repeat this experiment 1000 times and aggregate the results. We find that our approach provides accurate values 59\% times when we conduct 200 simulations compared to the average-based approach (41\% times). We run 1000 simulations and find out that our approach has a high likelihood (56.6\%) of giving accurate values compared to the other approach (43.4\%).

\vspace{-3mm}

\begin{figure}[htbp]
\begin{center}
\vspace{-2mm}
\includegraphics[width=.85\columnwidth,height=5cm]{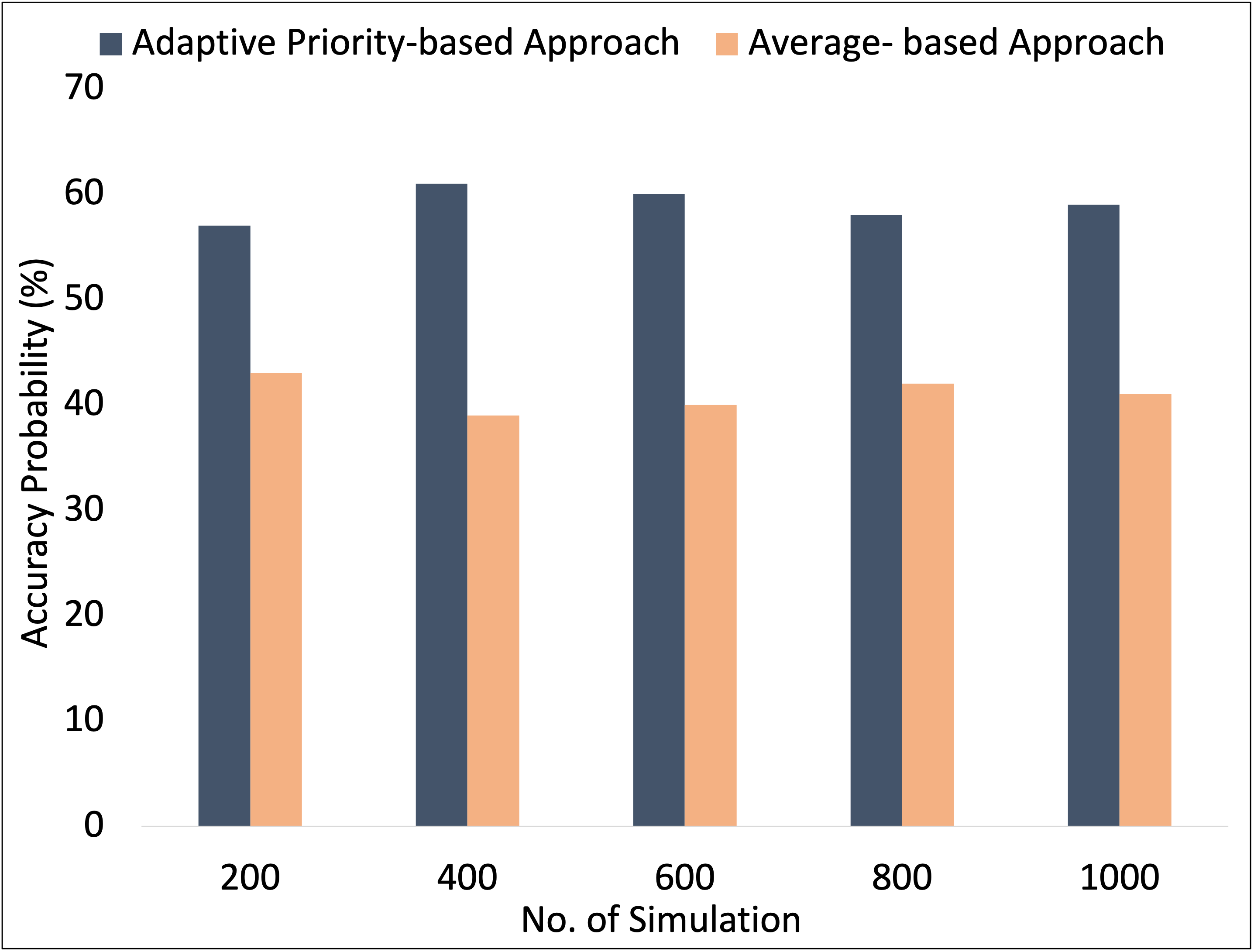}
\vspace{-2mm}
\caption{Accuracy probability in uniform distribution.}
\label{fig5}
\end{center}
\vspace{-6mm}
\end{figure}

The second set of experiments is conducted considering ground truth values generated from a uniform distribution. The results are shown in Fig. \ref{fig5}. We follow the same principle for determining values (i.e., min-max value) for uniform distribution as described in the above paragraph. We run the simulations 200 times, 400 times, 600 times, and so on. In total, we repeat this experiment 1000 times and aggregate the results. We find that our approach provides accurate values 57\% times when we conduct 200 simulations compared to the average-based approach (43\% times). When we run 400 simulations, there is a 61\% likelihood that our approach provides accurate values. We run 1000 simulations and find out that the proposed approach has a high likelihood (59\%) of providing accurate values than the average-based approach (41\%).

\begin{figure}[htbp]
\begin{center}
\vspace{-2mm}
\includegraphics[width=.85\columnwidth,height=5cm]{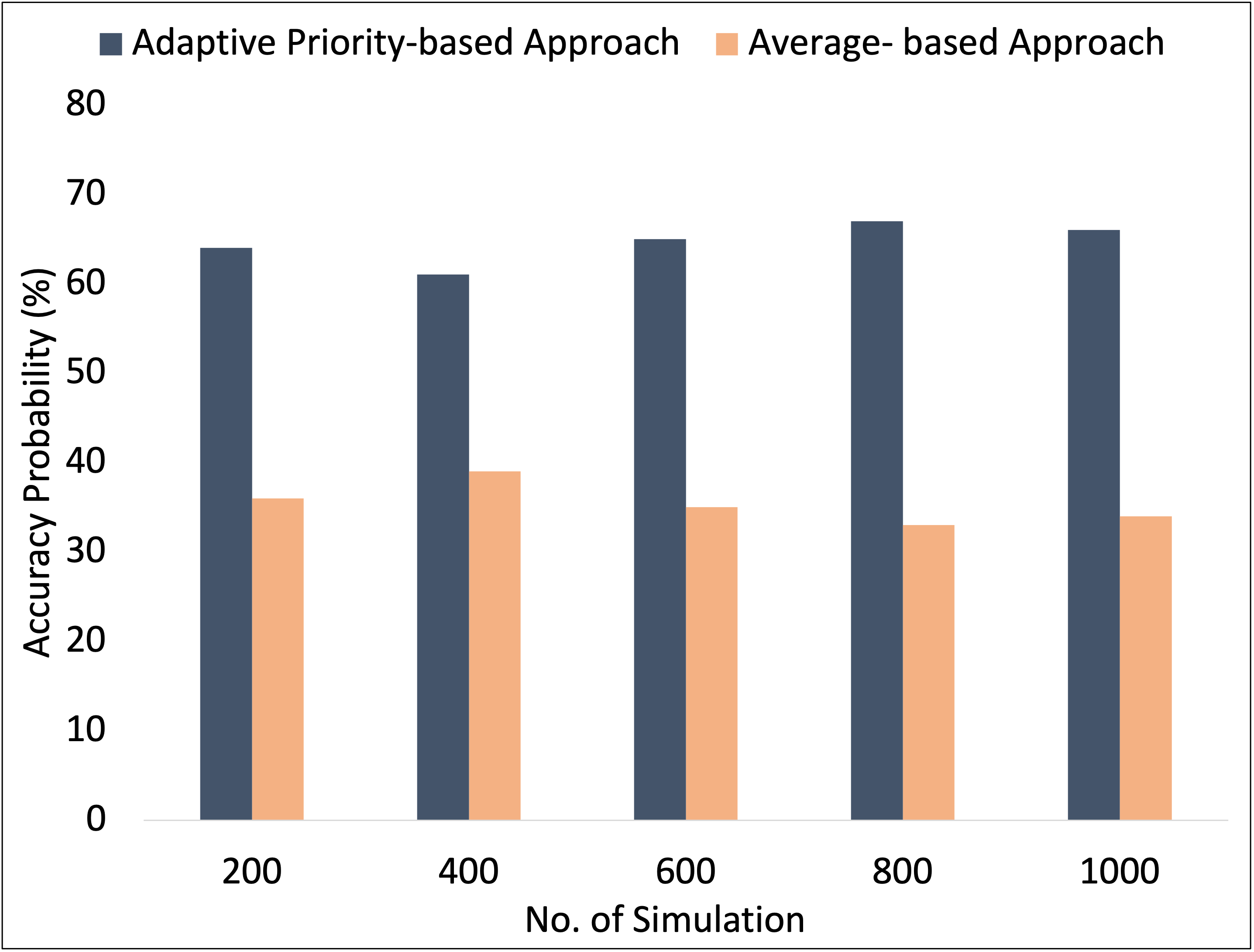}
\vspace{-2mm}
\caption{Accuracy probability in triangular distribution.}
\vspace{-3mm}
\label{fig6}
\end{center}
\end{figure}

The third set of experiments is conducted considering ground truth values generated from triangular distribution. The results are shown in Fig. \ref{fig6}. A probability distribution function is required when generating data in triangular distribution. We provide minimum value, maximum value, and mode value for the distribution function. In order to get these values, we search the previous history to generate preference, and we use the current contextual data as well. We run the simulations 200 times, 400 times, 600 times, and so on. We repeat this experiment 1000 times and aggregate the results. We find that our approach provides accurate values 64\% times when we run 200 simulations compared to the average-based approach (36\% times). When we run 400 simulations, there is a 61\% likelihood that our approach gives values close to ground truth. In total, we run 1000 simulations and find out that the proposed adaptive priority-based approach has a high likelihood (64.6\%) of giving accurate values than the average-based approach (35.4\%).

\begin{figure}[htbp]
\begin{center}
\vspace{-2mm}
\includegraphics[width=.85\columnwidth,height=5cm]{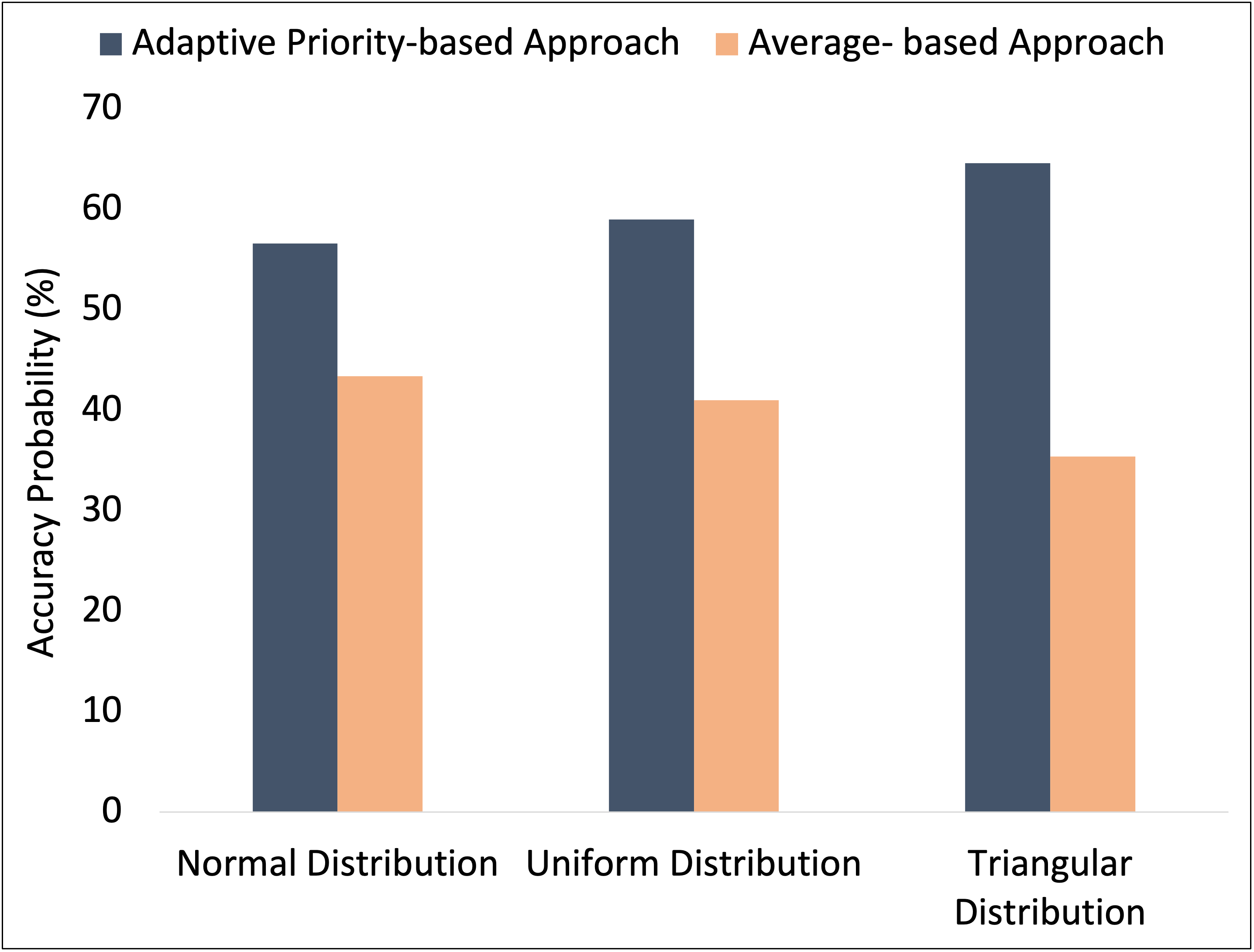}
\vspace{-2mm}
\caption{Accuracy probability in all distributions.}
\vspace{-3mm}
\label{fig7}
\end{center}
\vspace{-3mm}
\end{figure}

After that, we calculate the overall performance of these two approaches by averaging all the experimental results and find that our proposed adaptive priority-based approach performs better than the other existing average-based approach (Fig. \ref{fig7}). It means there is a high likelihood that our proposed approach will provide values (i.e., AC temperature, light illumination) close to the ground truth values. Monte Carlo methods are used when there is uncertainty, and there is no deterministic result. The overall idea of testing the conflict resolution framework is probabilistic, and the Monte Carlo simulation is an excellent fit to evaluate our proposed model.

\section{Conclusion and Future Work}

We propose a novel approach for conflict resolution of IoT services by assigning priority to the residents based on contextual factors such as age, illness, visual impairment, and hearing impairment. The proposed adaptive priority model is developed using the analytic hierarchy approach as it allows the contextual factors to be arranged in a hierarchic structure. The effectiveness of the proposed conflict resolution approach are tested with another existing approach. The future direction is to recommend alternative services to make residents' lives more comfortable and convenient.

\vspace{-2mm}



\def\IEEEbibitemsep{0pt plus 0.1pt}
\bibliographystyle{IEEEtran}
\bibliography{ConfRes}



%




\end{document}